\documentclass[conference,10pt,twocolumn]{IEEEtran}

\usepackage{amsmath,amssymb,amsfonts}
\usepackage{bm}
\usepackage{tabularx}
\usepackage{color}
\usepackage{cite}
\usepackage{algorithm,algorithmic}
\usepackage{caption}
\usepackage{subcaption}
\usepackage{lettrine}
\usepackage{float}
\usepackage[export]{adjustbox}
\usepackage{graphicx}
\usepackage{booktabs}
\usepackage[dvips]{epsfig}
\usepackage{boxedminipage}

\usepackage{scalerel}
\usepackage{tikz}
\usetikzlibrary{svg.path}

\definecolor{orcidlogocol}{HTML}{A6CE39}
\tikzset{
    orcidlogo/.pic={
        \fill[orcidlogocol] svg{M256,128c0,70.7-57.3,128-128,128C57.3,256,0,198.7,0,128C0,57.3,57.3,0,128,0C198.7,0,256,57.3,256,128z};
        \fill[white] svg{M86.3,186.2H70.9V79.1h15.4v48.4V186.2z}
        svg{M108.9,79.1h41.6c39.6,0,57,28.3,57,53.6c0,27.5-21.5,53.6-56.8,53.6h-41.8V79.1z M124.3,172.4h24.5c34.9,0,42.9-26.5,42.9-39.7c0-21.5-13.7-39.7-43.7-39.7h-23.7V172.4z}
        svg{M88.7,56.8c0,5.5-4.5,10.1-10.1,10.1c-5.6,0-10.1-4.6-10.1-10.1c0-5.6,4.5-10.1,10.1-10.1C84.2,46.7,88.7,51.3,88.7,56.8z};
    }
}

\newcommand\orcidicon[1]{\href{https://orcid.org/#1}{\mbox{\scalerel*{
                \begin{tikzpicture}[yscale=-1,transform shape]
                \pic{orcidlogo};
                \end{tikzpicture}
            }{|}}}}

\usepackage[bookmarks=false]{hyperref}

\DeclareCaptionLabelSeparator{periodspace}{.\quad}
\DeclareMathOperator*{\argmin}{argmin}
\IEEEoverridecommandlockouts

\begin{document}
\title{Atomic Norm Minimization-based Low-Overhead Channel Estimation for RIS-aided MIMO Systems 
\thanks{This research was supported by the MSIP (Ministry of Science, ICT and Future Planning), Korea, under the ITRC (Information Technology Research Center) support program (IITP-2021-2017-0-01637) supervised by the IITP (Institute for Information \& communications Technology Promotion).}}

\author{
\IEEEauthorblockN{Hyeonjin Chung and Sunwoo Kim \\ Department of Electronics and Computer Engineering, Hanyang University, Seoul, South Korea \\ E-mail: \{hyeonjingo,remero\}@hanyang.ac.kr}}
\maketitle

\begin{abstract}
Large beam training overhead has been considered as one of main issues in the channel estimation for reconfigurable intelligent surface (RIS)-aided systems.
In this paper, we propose an atomic norm minimization (ANM)-based low-overhead channel estimation for RIS-aided multiple-input-multiple-output (MIMO) systems.
When the number of beam training is reduced, some multipath signals may not be received during beam training, and this causes channel estimation failure.
To solve this issue, the width of beams created by RIS is widened to capture all multipath signals. 
Pilot signals received during beam training are compiled into one matrix to define the atomic norm of the channel for RIS-aided MIMO systems.
Simulation results show that the proposed algorithm outperforms other channel estimation algorithms.
\end{abstract}

% Note that keywords are not normally used for peer review papers.

\IEEEpeerreviewmaketitle

\section{Introduction}
Reconfigurable intelligent surface (RIS) has been received considerable attention owing to the property that it can programmably change the propagation characteristic of the signal~\cite{9140329}.
When it comes to millimeter-wave communications, RIS can make the signal bypass the blockage by reflecting the signal from the base station (BS) to the user equipment (UE)~\cite{9119122}. 
The advent of RIS has generated various new research, and one of them is the channel estimation for RIS-aided systems.
One of major concerns in the channel estimation for RIS-aided systems is that the RIS has been added to the link between BS and UE, and the addition of RIS causes the excessive beam training overhead for channel estimation. 

To address the issue on large overhead, the channel estimation algorithms based on compressive sensing (CS)~\cite{9103231,9354904,chen2019channel} and atomic norm minimization (ANM)~\cite{he2021channel} have been introduced, assuming that signal paths are sparse.
% In~\cite{chen2019channel}, the uplink channel estimation for RIS-aided multi-user multiple-input-multiple-output (MISO) systems is proposed, where the system in~\cite{chen2019channel} employs one BS equipped with multiple antennas and multiple UEs equipped with single antenna. 
Channel estimation algorithms for RIS-aided systems in~\cite{9103231,9354904,he2021channel} commonly estimate angle-of-departures (AoDs) and angle-of-arrivals (AoAs), then estimate channel gains to construct the channel. 
Algorithms in~\cite{9103231,9354904} exploit CS for AoD/AoA estimation, but in this case, a grid-mismatch limits the estimation accuracy~\cite{5710590}.
In~\cite{he2021channel}, the grid-mismatch is handled by ANM, however, the channel estimation becomes inaccurate when AoAs and AoDs are closely separated.
The channel estimation for RIS-aided multi-user systems is proposed in~\cite{chen2019channel}, where the algorithm in~\cite{chen2019channel} jointly estimates multiple channels by using CS-based multi-user joint channel estimator.

Although there have been various studies on the low-overhead channel estimation, the problem induced by the shortage of training beams and how it affects the channel estimation have not been discussed properly.
% As \cite{9129778} pointed out, beams created by RIS have to explore all directions during the beam training. Otherwise, there is a possibility of the signal reception failure.
In this paper, we propose an ANM-based low-overhead channel estimation for RIS-aided multiple-input-multiple-output (MIMO) systems.
When the beam training is reduced, an erroneous multipath reception may occur and induces the channel estimation failure.
To tackle this issue, a training beamwidth adaptation is proposed to widen the beamwidth when there is less beam training.
Then, the atomic norm of the channel for RIS-aided MIMO systems is defined, where defining the atomic norm is feasible when pilot signals received during beam training are compiled in a specified manner. 
A detailed explanation on the proposed algorithm is presented in following sections.

$\textit{Notations:}$ We use lower-case and upper-case bold characters to respectively represent vectors and matrices throughout this paper. $(\cdot)^{T}$, $(\cdot)^{H}$, and $(\cdot)^{*}$ respectively denote transpose, conjugate transpose, and complex conjugation.
$\textrm{Tr}(\cdot)$ denotes the trace of a matrix, and $\textrm{diag}(\cdot)$ denotes the diagonal matrix whose diagonal entries equal to entries of given vector. $\textrm{vec}(\cdot)$ denotes vectorization of given matrix.
$\lVert \cdot \rVert_{\textrm{F}}$ denotes Frobenius norm. 
The curled inequality symbol $\succeq$ denotes matrix inequality. If $\mathbf{A} \succeq \mathbf{B}$, a matrix $\mathbf{A}-\mathbf{B}$ is positive semidefinite.
$\otimes$ and $\diamond$ respectively denote Kronecker product and Khatri-Rao product.
$\mathbf{I}_{N}$ denotes a $N \times N$ identity matrix.

\section{Channel and Signal Model}\label{system model}
We consider a downlink RIS-aided MIMO system, which means that a base station (BS) transmits the signal to a RIS and the RIS bounces back the signal to an user equipment (UE).
The BS, the RIS, and the UE are equipped with $M_{\textrm{B}}$, $M_{\textrm{R}}$, and $M_{\textrm{U}}$ antennas respectively. Here, antenna arrays that BS, RIS, and UE use are uniform linear arrays (ULAs) with half-lambda spacing. 
In this paper, the BS and the UE employ full-complexity hybrid beamforming structure~\cite{hybrid}, where the BS and the UE are respectively equipped with $N_{\textrm{B}}$ and $N_{\textrm{U}}$ RF chains. 

\begin{figure}[!t]
    \begin{center}
    \includegraphics[width=0.9\columnwidth]{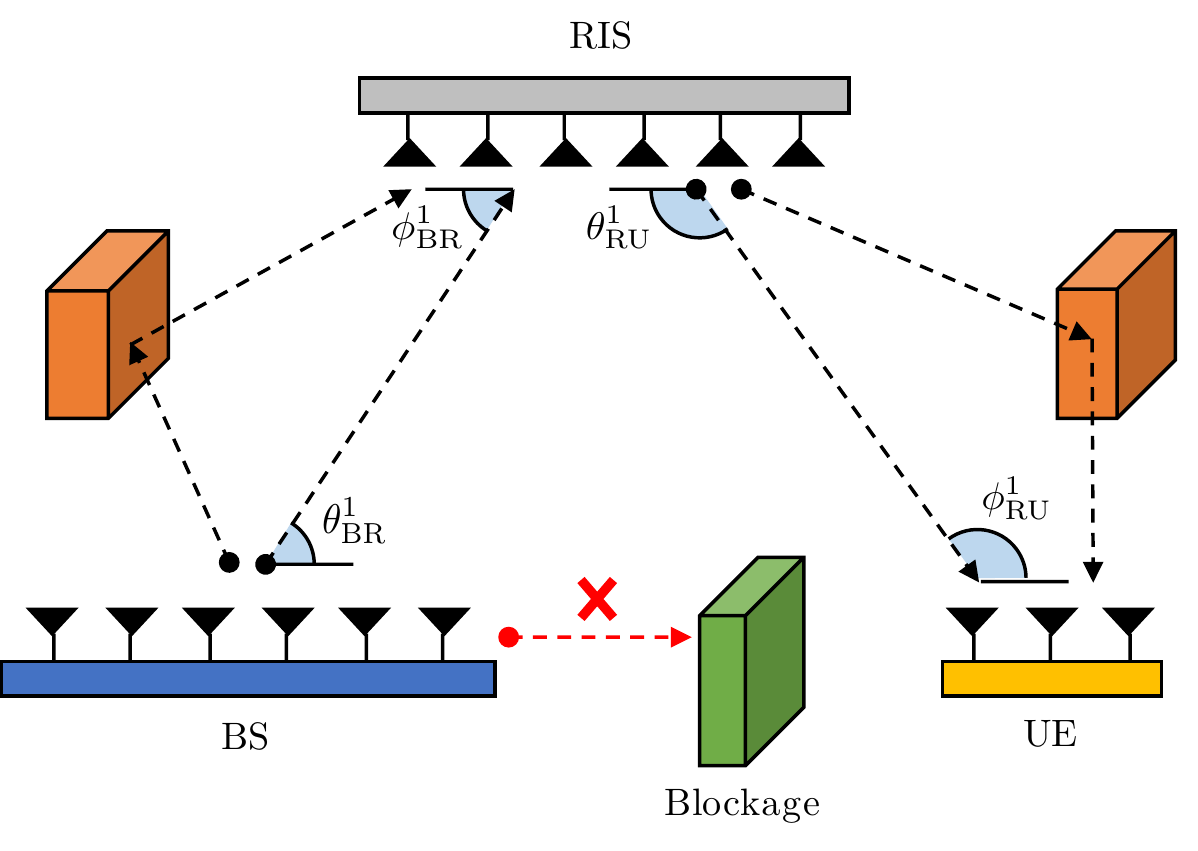}
    \caption{A scheme of RIS-aided MIMO system. The signal path between BS and UE is blocked.}
    \label{RIS}
    \end{center}
\end{figure}

The steering vector of the ULA with half-lambda spacing, $\mathbf{a}(\theta)$ is
\begin{equation}\label{steer_vec}
    \mathbf{a}(\theta) = [1,e^{j\pi\cos\theta},\ldots,e^{j\pi(M-1)\cos\theta}]^T \in \mathbb{C}^{M\times 1},
\end{equation}
where $\theta$ denotes the steering direction, and $M$ denotes the number of antennas.
A scheme of RIS-aided MIMO system considered in this paper is given in Fig.~\ref{RIS}.
Assuming all signal paths between BS and UE are blocked, a channel for RIS-aided MIMO system can be represented as a cascade of two separate channels: BS-to-RIS channel and RIS-to-UE channel.
The BS-to-RIS channel $\mathbf{H}_{\textrm{BR}}$ can be given by
\begin{equation}\label{HBR}
    \begin{split}
    \mathbf{H}_{\textrm{BR}} &=\sum_{l=1}^{L_{\textrm{BR}}} \alpha_{\textrm{BR}}^{l} \mathbf{a}(\phi_{\textrm{BR}}^{l}) \mathbf{a}(\theta_{\textrm{BR}}^{l})^{H} \\
    &= \mathbf{A}(\bm{\phi}_{\textrm{BR}}) \textrm{diag}(\bm{\rho}_{\textrm{BR}}) \mathbf{A}(\bm{\theta}_{\textrm{BR}})^{H} \in \mathbb{C}^{M_{\textrm{R}} \times M_{\textrm{B}}},
    \end{split}
\end{equation}
where $L_{\textrm{BR}}$ denotes the number of signal paths between BS and RIS. $\alpha_{\textrm{BR}}^{l}$, $\phi_{\textrm{BR}}^{l}$, and $\theta_{\textrm{BR}}^{l}$ respectively denote the channel gain, the BS-to-RIS AoA, and the BS-to-RIS AoD of the $l$-th signal path. 
$\bm{\phi}_{\textrm{BR}}= \{ {\phi}^{1}_{\textrm{BR}},\ldots,{\phi}^{L_{\textrm{BR}}}_{\textrm{BR}} \}$ and $\bm{\theta}_{\textrm{BR}}= \{ {\theta}^{1}_{\textrm{BR}},\ldots,{\theta}^{L_{\textrm{BR}}}_{\textrm{BR}} \}$.
$\mathbf{A}(\bm{\phi}_{\textrm{BR}})=[\mathbf{a}(\phi_{\textrm{BR}}^{1}),\ldots,\mathbf{a}(\phi_{\textrm{BR}}^{L_{\textrm{BR}}}) ] \in \mathbb{C}^{M_{\textrm{R}} \times L_{\textrm{BR}}}$, $\mathbf{A}(\bm{\theta}_{\textrm{BR}})=[\mathbf{a}(\theta_{\textrm{BR}}^{1}),\ldots,\mathbf{a}(\theta_{\textrm{BR}}^{L_{\textrm{BR}}}) ] \in \mathbb{C}^{M_{\textrm{B}} \times L_{\textrm{BR}}}$, and $\bm{\rho}_{\textrm{BR}}=[\alpha_{\textrm{BR}}^{1},\ldots,\alpha_{\textrm{BR}}^{L_{\textrm{BR}}}]^{T}$. 
The RIS-to-UE channel $\mathbf{H}_{\textrm{RU}}$ can be given by
\begin{equation}\label{HRU}
    \begin{split}
    \mathbf{H}_{\textrm{RU}} &=\sum_{l=1}^{L_{\textrm{RU}}} \alpha_{\textrm{RU}}^{l} \mathbf{a}(\phi_{\textrm{RU}}^{l}) \mathbf{a}(\theta_{\textrm{RU}}^{l})^{H} \\
    &= \mathbf{A}(\bm{\phi}_{\textrm{RU}}) \textrm{diag}(\bm{\rho}_{\textrm{RU}}) \mathbf{A}(\bm{\theta}_{\textrm{RU}})^{H} \in \mathbb{C}^{M_{\textrm{U}} \times M_{\textrm{R}}},
    \end{split}
\end{equation}
where $L_{\textrm{RU}}$ denotes the number of signal paths between RIS and UE. $\alpha_{\textrm{RU}}^{l}$, $\phi_{\textrm{RU}}^{l}$, and $\theta_{\textrm{RU}}^{l}$ respectively denote the channel gain, the RIS-to-UE AoA, and the RIS-to-UE AoD of the $l$-th signal path.
$\bm{\phi}_{\textrm{RU}}= \{ {\phi}^{1}_{\textrm{RU}},\ldots,{\phi}^{L_{\textrm{RU}}}_{\textrm{RU}} \}$ and $\bm{\theta}_{\textrm{RU}}= \{ {\theta}^{1}_{\textrm{RU}},\ldots,{\theta}^{L_{\textrm{RU}}}_{\textrm{RU}} \}$.
$\mathbf{A}(\bm{\phi}_{\textrm{RU}})=[\mathbf{a}(\phi_{\textrm{RU}}^{1}),\ldots,\mathbf{a}(\phi_{\textrm{RU}}^{L_{\textrm{RU}}}) ] \in \mathbb{C}^{M_{\textrm{U}} \times L_{\textrm{RU}}}$, $\mathbf{A}(\bm{\theta}_{\textrm{RU}})=[\mathbf{a}(\theta_{\textrm{RU}}^{1}),\ldots,\mathbf{a}(\theta_{\textrm{RU}}^{L_{\textrm{RU}}}) ] \in \mathbb{C}^{M_{\textrm{R}} \times L_{\textrm{RU}}}$, and $\bm{\rho}_{\textrm{RU}}=[\alpha_{\textrm{RU}}^{1},\ldots,\alpha_{\textrm{RU}}^{L_{\textrm{RU}}}]^{T}$.

A RIS control matrix $\bm{\Omega}$ can be given by
\begin{equation}\label{Omega}
    \bm{\Omega}=
    \begin{bmatrix}
    \beta_{1} e^{j \vartheta_{1}} & 0 & \ldots & 0 \\
    0 & \beta_{2} e^{j \vartheta_{2}} & \ldots & 0 \\
    \vdots & \vdots & \ddots & \vdots \\
    0 & 0 & \ldots & \beta_{M_{\textrm{R}}} e^{j \vartheta_{M_{\textrm{R}}}}
    \end{bmatrix} \in \mathbb{C}^{M_{\textrm{R}} \times M_{\textrm{R}}},
\end{equation}
where $\beta_{m}$ and $\vartheta_{m}$ respectively denote a reflection coefficient and a phase shift of the $m$-th antenna in RIS.
$\vartheta_{m} \in [0,2\pi)$ and $\beta_{m}$ can be either $0$ or $1$, where $0$ and $1$ respectively denotes the deactivation and the activation of the $m$-th antenna in the RIS.  
For a convenient representation of $\bm{\Omega}$, a RIS control vector $\bm{\omega}$ is defined as
\begin{equation}\label{vecOmega}
    \bm{\omega}=\left[\beta_{1} e^{j \vartheta_{1}},\beta_{2} e^{j \vartheta_{2}},\ldots,\beta_{M_{\textrm{R}}} e^{j \vartheta_{M_{\textrm{R}}}} \right]^{T} \in \mathbb{C}^{M_{\textrm{R}} \times 1}.
\end{equation}
Note that $\bm{\Omega}=\textrm{diag}(\bm{\omega})$. The cascaded channel for RIS-aided MIMO system, $\mathbf{H}$ can be given by
\begin{equation}\label{Cascade}
    \mathbf{H}=\mathbf{H}_{\textrm{RU}} \bm{\Omega} \mathbf{H}_{\textrm{BR}} \in \mathbb{C}^{M_{\textrm{U}} \times M_{\textrm{B}}}.
\end{equation}

\begin{figure}[!t]
    \includegraphics[width=1\columnwidth]{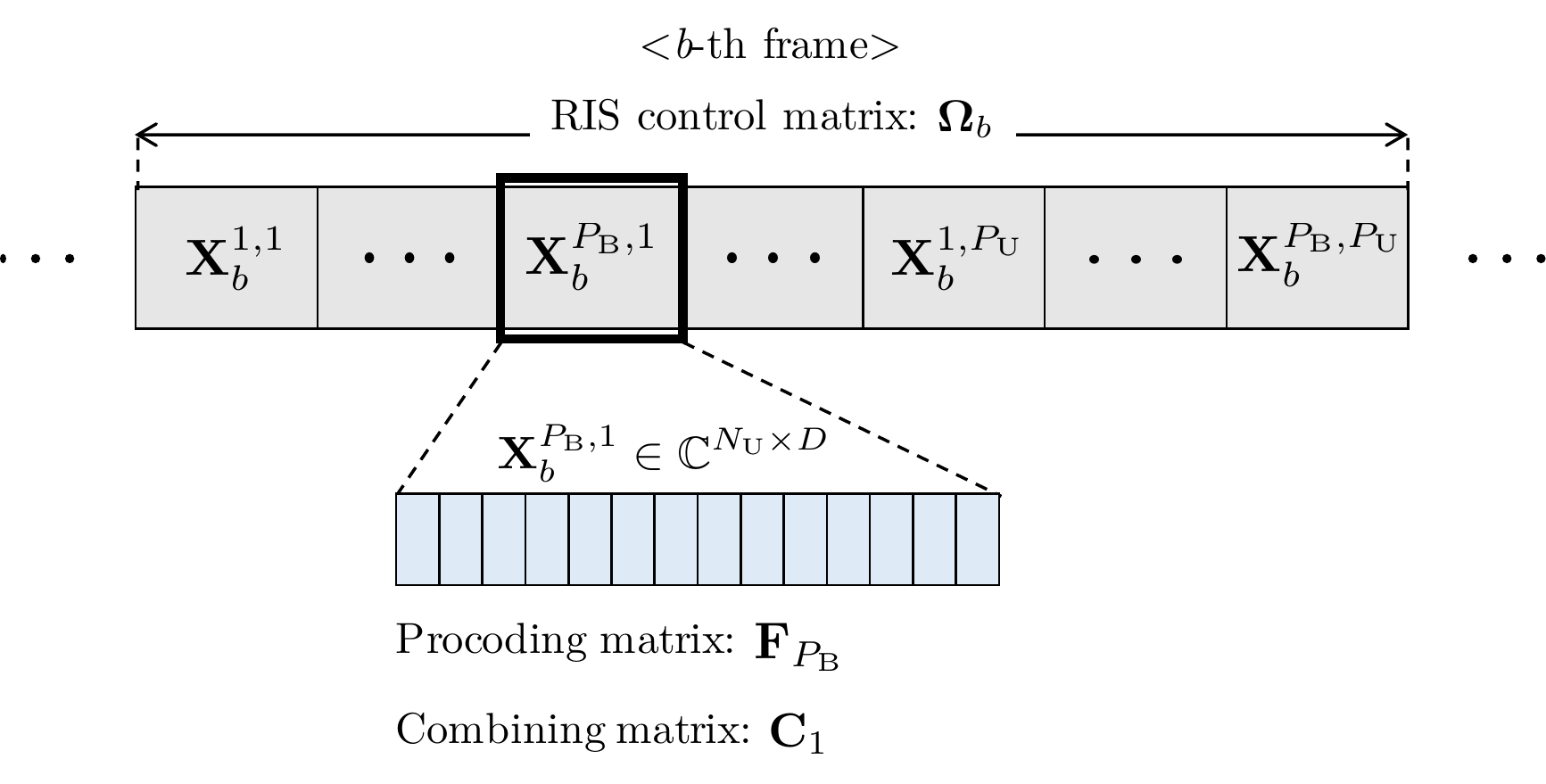}
    \caption{A frame structure for beam training procedure in RIS-aided MIMO communication.}
    \label{Fig1}
\end{figure}

A frame structure for beam training procedure in RIS-aided MIMO systems is depicted in Fig.~\ref{Fig1}. 
To simplify notations, $M_{\textrm{B}}/N_{\textrm{B}}$ and $M_{\textrm{U}}/N_{\textrm{U}}$ are respectively defined as $P_{\textrm{B}}$ and $P_{\textrm{U}}$.
There are $P_{\textrm{B}}$ precoding matrices and $P_{\textrm{U}}$ combining matrices, and $\mathbf{F}_{i} \in \mathbb{C}^{M_{\textrm{B}} \times N_{\textrm{B}}}$ and $\mathbf{C}_{j} \in \mathbb{C}^{M_{\textrm{U}} \times N_{\textrm{U}}}$ respectively denote the $i$-th precoding matrix and the $j$-th combining matrix. 
The RIS control matrix changes frame by frame, and the BS and the UE perform a total $P_{\textrm{B}}P_{\textrm{U}}$ beam training at each frame. 
Letting $B$ denotes the number of frames, the total number of beam training $P$ equals to $B P_{\textrm{B}}P_{\textrm{U}}$. 
$\mathbf{X}^{i,j}_{b}$, a received pilot signal at the $b$-th frame which uses the $i$-th precoding matrix and the $j$-th combining matrix, can be given by 
\begin{equation}\label{rxsignal1}
    \mathbf{X}^{i,j}_{b}=\mathbf{C}_{j}^{H} \mathbf{H}_{\textrm{RU}} \bm{\Omega}_{b} \mathbf{H}_{\textrm{BR}} \mathbf{F}_{i} \mathbf{S} + \mathbf{N}^{i,j}_{b} \in \mathbb{C}^{N_{\textrm{U}} \times D},
\end{equation}
where $\bm{\Omega}_{b}$ denotes the RIS control matrix at the $b$-th frame, and $D$ denotes the number of signal samples per one beam training. $\mathbf{S}=\left[\mathbf{s}_{1},\ldots,\mathbf{s}_{N_{\textrm{B}}} \right]^{T} \in \mathbb{C}^{N_{\textrm{B}} \times D}$, where $\mathbf{s}_{n}$ is the $n$-th unit-energy pilot signal that satisfies $\mathbf{s}^{H}_{n}\mathbf{s}_{n}/D=1$. 
Note that pilot signals are orthogonal to each other so that $\mathbf{S}\mathbf{S}^{H}/D=\mathbf{I}_{N_{\textrm{B}}}$. 
$\mathbf{N}^{i,j}_{b} \in \mathbb{C}^{N_{\textrm{U}} \times D}$ is a noise matrix whose entries follow a circularly-symmetric complex Gaussian distribution. Here, the mean and the covariance are $0$ and $\sigma^{2}$ respectively. 

\begin{figure*}[!t]
    \begin{center}
    \captionsetup[subfigure]{justification=centering}
    \begin{subfigure}[t]{0.82\columnwidth}
        \includegraphics[width=\columnwidth,center]{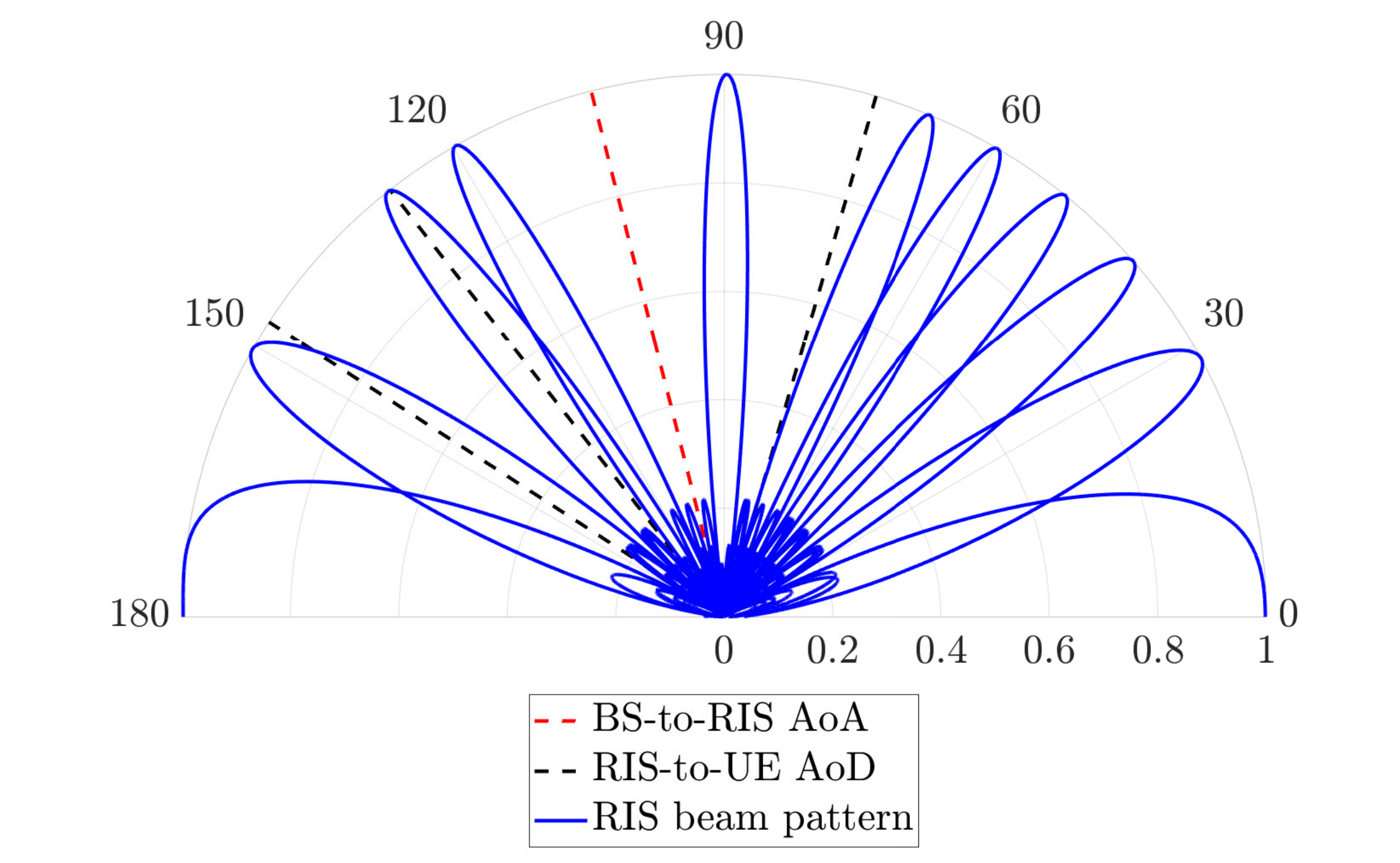}
        \caption{Radiation patterns without training beamwidth adaptation}\label{2a}
    \end{subfigure}
    \begin{subfigure}[t]{0.82\columnwidth}
        \includegraphics[width=\columnwidth,center]{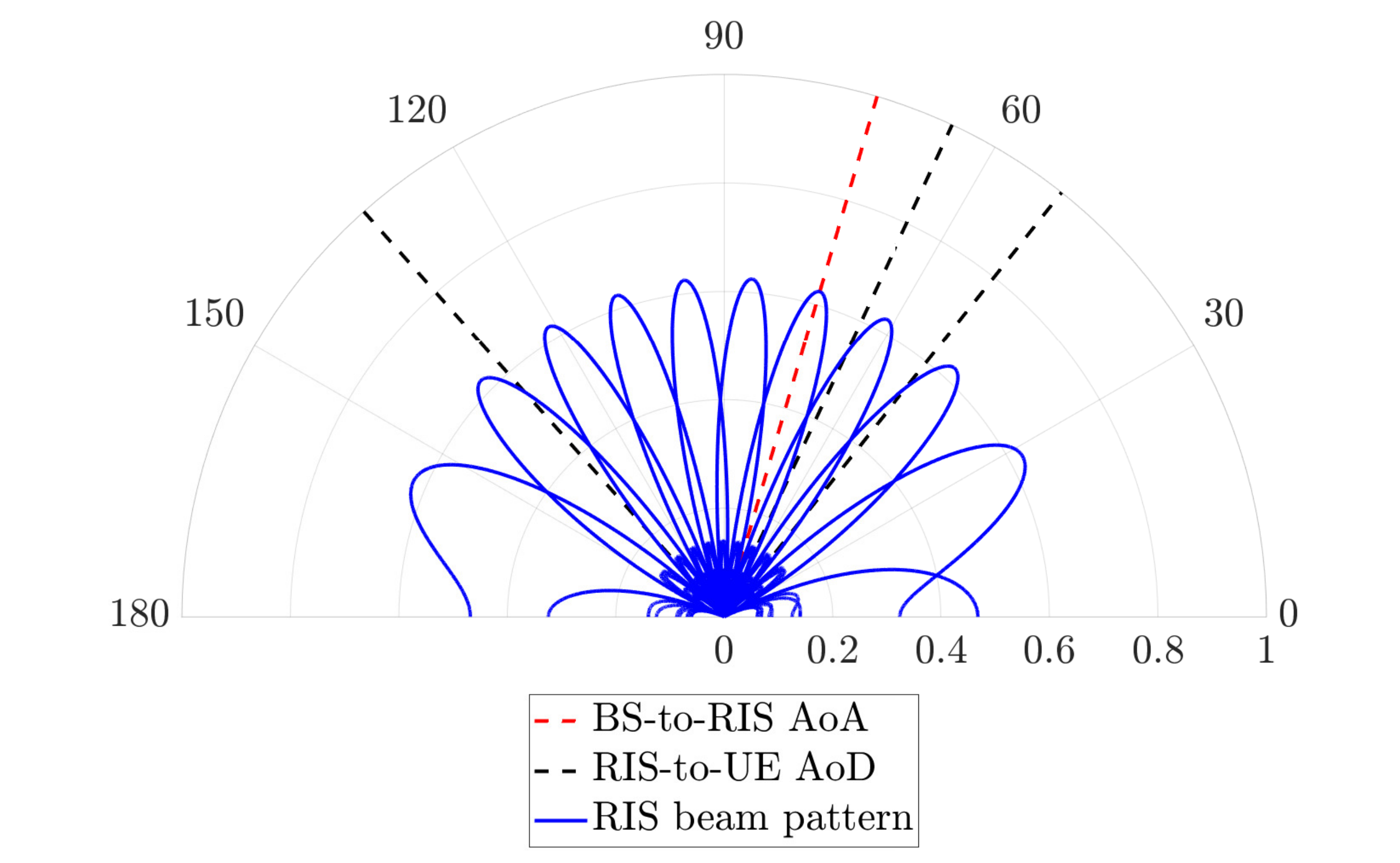}
        \caption{Radiation patterns with training beamwidth adaptation}\label{2b}
    \end{subfigure}
    \caption{Radiation patterns of beams created by RIS when $L_{\textrm{BR}}=1$, $L_{\textrm{RU}}=3$, $M_{\textrm{R}}=16$, and $B=10$. For successful multipath signal reception, every RIS-to-UE AoD has to be captured within one of beams.} 
    \label{all}
    \end{center}
\end{figure*}

After receiving $\mathbf{X}^{i,j}_{b}$, $\mathbf{S}^{H}$ are multiplied to $\mathbf{X}^{i,j}_{b}$ to filter the noise. 
We define this filtered signal $\mathbf{Y}^{i,j}_{b}$ as   
\begin{equation}\label{rxsignal2}
    \mathbf{Y}^{i,j}_{b}=\frac{\mathbf{X}^{i,j}_{b}\mathbf{S}^{H}}{D}=\mathbf{C}_{j}^{H} \mathbf{H}_{\textrm{RU}} \bm{\Omega}_{b} \mathbf{H}_{\textrm{BR}} \mathbf{F}_{i} + \frac{\mathbf{N}^{i,j}_{b}\mathbf{S}^{H}}{D} \in \mathbb{C}^{N_{\textrm{U}} \times N_{\textrm{B}}}.
\end{equation}
A total $P_{\textrm{B}}P_{\textrm{U}}$ filtered signals received at the $b$-th frame is organized as
\begin{equation}\label{merged}
    \mathbf{Y}_{b}= 
    \begin{bmatrix}
    \mathbf{Y}^{1,1}_{b} & \mathbf{Y}^{2,1}_{b} & \cdots & \mathbf{Y}^{P_{\textrm{B}},1}_{b} \\
    \mathbf{Y}^{1,2}_{b} & \mathbf{Y}^{2,2}_{b} & \cdots & \mathbf{Y}^{P_{\textrm{B}},2}_{b} \\
    \vdots & \vdots & \cdots & \vdots \\
    \mathbf{Y}^{1,P_{\textrm{U}}}_{b} & \mathbf{Y}^{2,P_{\textrm{U}}}_{b} & \cdots & \mathbf{Y}^{P_{\textrm{B}},P_{\textrm{U}}}_{b}
    \end{bmatrix}
    \in \mathbb{C}^{M_{\textrm{U}} \times M_{\textrm{B}}},
\end{equation}
where $\mathbf{Y}_{b}$ is a compilation of all filtered signals at the $b$-th frame. $\mathbf{Y}_{b}$ can also be represented as
\begin{equation}\label{merged2}
    \mathbf{Y}_{b}=\mathbf{C}^{H} \mathbf{H}_{\textrm{RU}} \bm{\Omega}_{b} \mathbf{H}_{\textrm{BR}} \mathbf{F} + \mathbf{V}_{b},
\end{equation}
where $\mathbf{F} \in \mathbb{C}^{M_{\textrm{B}} \times M_{\textrm{B}}}$ and $\mathbf{C} \in \mathbb{C}^{M_{\textrm{U}} \times M_{\textrm{U}}}$ respectively denote a full-rank precoding matrix and a full-rank combining matrix. 
$\mathbf{V}_{b} \in \mathbb{C}^{M_{\textrm{U}} \times M_{\textrm{B}}}$ is a matrix that represents the remaining noise.

\section{The Proposed Low-Overhead Channel Estimation for RIS-aided MIMO Systems}
\subsection{Compilation of Filtered Signals and Its Representation via Kronecker Product and Khatri-Rao Product}
To define the atomic norm of the channel, we compile filtered signals and present an organized representation of compiled filtered signals. 
For the compilation, two following properties are employed.
\begin{itemize}
	\item Property 1: $\textrm{vec}\left(\mathbf{A \textrm{diag}(\mathbf{b}) C} \right)=(\mathbf{C}^{T} \diamond \mathbf{A})\mathbf{b}.$
	\item Property 2: $(\mathbf{AB} \diamond \mathbf{CD})=(\mathbf{A} \otimes \mathbf{C})(\mathbf{B} \diamond \mathbf{D}).$
\end{itemize}
Definitions and properties of Kronecker product and Khatri-Rao product are well-explained in~\cite{KR}.
Letting a lengthy column vector $\mathbf{y}_{b}$ equals $\textrm{vec}(\mathbf{Y}_{b})$, $\mathbf{y}_{b}$ can be represented as follows by using Property 1.
\begin{equation}
    \mathbf{y}_{b}=\textrm{vec}(\mathbf{Y}_{b})=\left(\mathbf{F}^{T} \mathbf{H}^{T}_{\textrm{BR}} \diamond \mathbf{C}^{H} \mathbf{H}_{\textrm{RU}}\right) \bm{\omega}_{b}  + \mathbf{v}_{b} \in \mathbb{C}^{M_{\textrm{B}}M_{\textrm{U}} \times 1},
\end{equation}
where $\bm{\omega}_{b}$ denotes the RIS control vector at the $b$-th frame, and $\mathbf{v}_{b}=\textrm{vec}(\mathbf{V}_{b})$.

Then, we form a matrix $\bm{\mathcal{Y}}$ which is constructed by stacking $\mathbf{y}_{b}$ for $b=1,\ldots,B$ as follows.
\begin{equation}
    \begin{split}
    \bm{\mathcal{Y}}&=\left[\mathbf{y}_{1},\ldots,\mathbf{y}_{B} \right]\\&=\left(\mathbf{F}^{T} \mathbf{H}^{T}_{\textrm{BR}} \diamond \mathbf{C}^{H} \mathbf{H}_{\textrm{RU}}\right) \mathbf{W}  + \bm{\mathcal{V}} \in \mathbb{C}^{M_{\textrm{B}}M_{\textrm{U}} \times B},
    \end{split}
\end{equation}
where $\mathbf{W}=[\bm{\omega}_{1},\ldots,\bm{\omega}_{B}] \in \mathbb{C}^{M_{\textrm{R}} \times B}$ and $\bm{\mathcal{V}}=[\mathbf{v}_{1},\ldots,\mathbf{v}_{B}] \in \mathbb{C}^{M_{\textrm{B}}M_{\textrm{U}} \times B}$. 
By using Property 2, $\bm{\mathcal{Y}}$ can be also represented as
\begin{equation}
    \bm{\mathcal{Y}}=\left(\mathbf{F}^{T} \otimes \mathbf{C}^{H}\right) \left( \mathbf{H}^{T}_{\textrm{BR}} \diamond  \mathbf{H}_{\textrm{RU}} \right) \mathbf{W}  + \bm{\mathcal{V}}.
\end{equation}
Here, $\mathbf{H}^{T}_{\textrm{BR}} \diamond  \mathbf{H}_{\textrm{RU}} \in \mathbb{C}^{M_{\textrm{B}}M_{\textrm{U}} \times M_{\textrm{R}}}$ contains a channel information that is independent of $\mathbf{F}$, $\mathbf{C}$, and $\mathbf{W}$. 
Once $\mathbf{H}^{T}_{\textrm{BR}} \diamond  \mathbf{H}_{\textrm{RU}}$ is successfully estimated, the optimal RIS control matrix that maximizes SNR can be derived by conducting singular value decomposition (SVD) to $\mathbf{H}^{T}_{\textrm{BR}} \diamond  \mathbf{H}_{\textrm{RU}}$~\cite{he2021channel}. 
Throughout this paper, we define $\mathbf{H}^{T}_{\textrm{BR}} \diamond  \mathbf{H}_{\textrm{RU}}$ as an \textit{effective channel} $\mathbf{H}_{\textrm{eff}}$, which is a goal of channel estimation for RIS-aided MIMO systems.
% Note that the approach of estimating effective channel has also been used in seminal works of channel estimation for RIS-aided systems~\cite{chen2019channel,9103231,9354904,he2021channel}.

With (\ref{HBR}), (\ref{HRU}), and Property 2, $\bm{\mathcal{Y}}$ can be fully unfolded as
\begin{equation}
    \begin{split}
    \bm{\mathcal{Y}}&=\left(\mathbf{F}^{T} \otimes \mathbf{C}^{H}\right) \left( \mathbf{A}(\bm{\theta}_{\textrm{BR}})^{*} \otimes \mathbf{A}(\bm{\phi}_{\textrm{RU}})  \right)\\ & \left(\textrm{diag}(\bm{\rho}_{\textrm{BR}}) \otimes \textrm{diag}(\bm{\rho}_{\textrm{RU}}) \right) \left( \mathbf{A}(\bm{\phi}_{\textrm{BR}})^{T} \diamond \mathbf{A}(\bm{\theta}_{\textrm{RU}})^{H} \right) \mathbf{W} + \bm{\mathcal{V}}.
    \end{split}
\end{equation}
To simplify $\mathbf{A}(\bm{\phi}_{\textrm{BR}})^{T} \diamond \mathbf{A}(\bm{\theta}_{\textrm{RU}})^{H} \in \mathbb{C}^{L_{\textrm{BR}}L_{\textrm{RU}} \times M_{\textrm{R}}}$, $\bm{\varphi}$ is defined as
\begin{equation}
    \begin{split}
    \bm{\varphi}= \{ \varphi_{i,j}&: \cos^{-1} (\cos \theta^{j}_{\textrm{RU}} - \cos \phi^{i}_{\textrm{BR}} ), \\ & i=1,\ldots,L_{\textrm{BR}},j=1,\ldots,L_{\textrm{RU}}\}.
    \end{split}
\end{equation}
Then, $\mathbf{A}(\bm{\phi}_{\textrm{BR}})^{T} \diamond \mathbf{A}(\bm{\theta}_{\textrm{RU}})^{H}$ can be rewritten as
\begin{equation}
    \begin{split}
    &\mathbf{A}(\bm{\phi}_{\textrm{BR}})^{T} \diamond \mathbf{A}(\bm{\theta}_{\textrm{RU}})^{H}=\mathbf{A}(\bm{\varphi})^{H}\\&=\left[\mathbf{a}(\varphi_{1,1}),\ldots,\mathbf{a}(\varphi_{1,L_{\textrm{RU}}}),\ldots,\mathbf{a}(\varphi_{L_{\textrm{BR}},1}),\ldots,\mathbf{a}(\varphi_{L_{\textrm{BR}},L_{\textrm{RU}}}) \right]^{H}.
    \end{split}
\end{equation}

\subsection{Robust Multipath Signal Reception When Using Fewer Beam Training via Training Beamwidth Adaptation}
One of general ways to control RIS is to use a discrete Fourier transform (DFT) matrix~\cite{5707050}. $N \times N$ DFT matrix $\bm{\Psi}_{N}$ can be given by
\begin{equation}
    \bm{\Psi}_{N}= 
    \begin{bmatrix}
    1 & 1 & \cdots & 1 \\
    1 & e^{j \frac{2\pi}{N}} & \cdots & e^{j \frac{2\pi (N-1)}{N}} \\
    1 & e^{j \frac{4\pi}{N}} & \cdots & e^{j \frac{4\pi (N-1)}{N}} \\
    \vdots & \vdots & \cdots & \vdots \\
    1 & e^{j 2\pi} & \cdots & e^{j 2\pi(N-1)}
    \end{bmatrix}
    \in \mathbb{C}^{N \times N}.
\end{equation}
For mainlobes of beams to cover the entire angular domain, $\mathbf{W}$ should be equal to $\bm{\Psi}_{M_{\textrm{R}}}$ so that $B=M_{\textrm{R}}$.
However, considering $P = B P_{\textrm{B}}P_{\textrm{U}}$, $B$ should be reduced in order to prevent the beam training overhead from getting excessively large.
In this subsection, we discuss on the erroneous multipath signal reception that occurs when $B < M_{\textrm{R}}$ and how to resolve it. 

The simplest way of determining RIS control vectors when $B < M_{\textrm{R}}$ is to select $B$ columns from $\bm{\Psi}_{M_{\textrm{R}}}$ as in~\cite{9354904}, but lack of beams can cause erroneous multipath signal reception as in Fig.~\ref{2a}.
Fig.~\ref{2a} shows the case of the erroneous multipath signal reception when one of RIS-to-UE signal paths does not fall onto mainlobes of $B$ beams. 
In this case, the channel estimation fails since signals from all paths are required for perfect channel estimation.  

To address the issue on the erroneous multipath signal reception, a training beamwidth adaptation is proposed to make multipath signal reception robust when $B<M_{\textrm{R}}$.
The beamwidth of RIS can be widened by deactivating the part of RIS, and $\mathbf{W}$ that contains $B$ widened beams can be given by
\begin{equation}
    \mathbf{W}=
    \begin{bmatrix}
    \bm{\Psi}_{B} \\ \mathbf{O}_{M_{\textrm{R}}-B,B}
    \end{bmatrix} \in \mathbb{C}^{M_\textrm{R} \times B},
\end{equation}
where $\mathbf{O}_{M,N}$ denotes a $M \times N$ zero matrix. Fig.~\ref{2b} shows beams that are widened by training beamwidth adaptation when $M_{\textrm{R}}=16$ and $B=10$. 
In Fig.~\ref{2b}, mainlobes of $B$ beams cover the entire angular domain so that every RIS-to-UE signal path is captured within one of $B$ beams, although the beam gain decreases. 

\subsection{Atomic Norm Minimization-based Channel Estimation for RIS-aided MIMO Systems}
The simplest approach to estimate the effective channel is to use least square (LS) estimator. 
If $\mathbf{F}^{T} \otimes \mathbf{C}^{H}$ and $\mathbf{W}$ are both full-rank matrices, the effective channel can be estimated via LS estimator as follows.
\begin{equation}
    \hat{\mathbf{H}}_{\textrm{eff}}^{\textrm{LS}}=\left(\mathbf{F}^{T} \otimes \mathbf{C}^{H}\right)^{-1} \bm{\mathcal{Y}} \mathbf{W}^{-1},
\end{equation}
where $\hat{\mathbf{H}}_{\textrm{eff}}^{\textrm{LS}}$ is the effective channel estimated by LS estimator.
The LS estimator requires at least $M_{\textrm{R}} P_{\textrm{B}}P_{\textrm{U}}$ beam training since $B$ should be larger than $M_{\textrm{R}}$ to make $\mathbf{W}$ a full-rank matrix.
$\mathbf{F}^{T} \otimes \mathbf{C}^{H}$ is full-rank since $\mathbf{F}$ and $\mathbf{C}$ are both full-rank matrices.
It is worth noting that the beam training overhead of LS estimator is generally large considering $M_{\textrm{R}}$ can be more than hundreds in practice~\cite{9086766}. 
On the other hand, the effective channel can be estimated by ANM even when $B<M_{\textrm{R}}$ if signal paths are sparse and all multipath signals are received properly.
In this subsection, an atomic norm of the effective channel is defined, where defining proper atomic norm leads to accurate channel estimation.  

\begin{figure}[!t]
    \includegraphics[width=1\columnwidth]{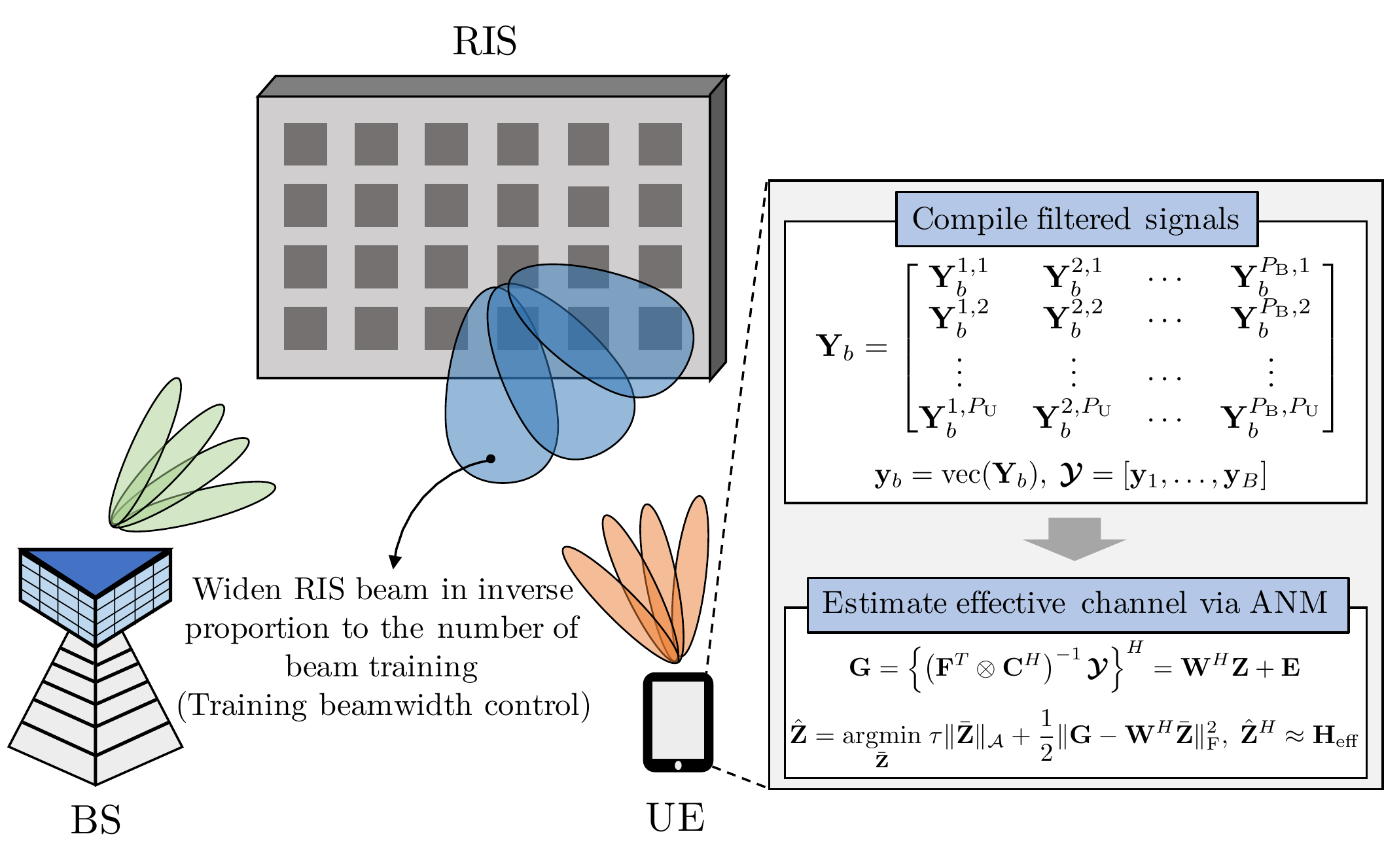}
    \caption{A summary of the proposed ANM-based low-overhead channel estimation for RIS-aided MIMO systems.}
    \label{Fig3}
\end{figure}

We assume the signal paths are sparse so that $B>L_{\textrm{BR}}L_{\textrm{RU}}$.
For simplification of equations and notations, we define $\mathbf{G}$ and $\mathbf{Z}$ as follows.
\begin{equation}
    \begin{split}
    \mathbf{G}=\left\{ \left(\mathbf{F}^{T} \otimes \mathbf{C}^{H}\right)^{-1} \bm{\mathcal{Y}} \right\}^{H}=\mathbf{W}^{H} \mathbf{Z} + \mathbf{E} \in \mathbf{C}^{B \times M_{\textrm{B}}M_{\textrm{U}}},  
    \end{split}
\end{equation}
\begin{equation}
    \begin{split}
    \mathbf{Z}&=\mathbf{H}_{\textrm{eff}}^{H}=\left( \mathbf{H}^{T}_{\textrm{BR}} \diamond  \mathbf{H}_{\textrm{RU}} \right)^{H}\\ &=\mathbf{A}(\bm{\varphi}) \left(\textrm{diag}(\bm{\rho}_{\textrm{BR}}) \otimes \textrm{diag}(\bm{\rho}_{\textrm{RU}}) \right)^{H} \left( \mathbf{A}(\bm{\theta}_{\textrm{BR}})^{*} \otimes \mathbf{A}(\bm{\phi}_{\textrm{RU}})  \right)^{H}.
    \end{split}
\end{equation}
Here, $\mathbf{E}=\{ \left(\mathbf{F}^{T} \otimes \mathbf{C}^{H}\right)^{-1} \bm{\mathcal{V}} \}^{H}$. 
For the estimation of $\mathbf{Z}$, an atomic set $\mathcal{A}$ is defined as follows.
\begin{equation}\label{atom}
    \mathcal{A}=\left\{\mathbf{a}(\theta)\mathbf{b}^{T} \in \mathbb{C}^{M_{\textrm{R}} \times M_{\textrm{B}}M_{\textrm{U}}}: 0^{\circ} < \theta < 180^{\circ}, \lVert \mathbf{b} \rVert_{2}=1  \right\},        
\end{equation}
where $\mathbf{a}(\theta)\mathbf{b}^{T}$ is defined as an atom, and $\mathbf{Z}$ can be represented as a linear combination of atoms.
Properties of the atom defined in~(\ref{atom}) and its atomic norm have been studied in seminal works of the ANM~\cite{9016105,7484756,7313018}.
The atomic norm of $\mathbf{Z}$, $\lVert \mathbf{Z} \rVert_{\mathcal{A}}$ can be represented by following SDP~\cite{9016105}: 
\begin{equation}
    \begin{split}   
    \lVert \mathbf{Z} \rVert_{\mathcal{A}}=&\min_{\mathbf{u},\mathbf{T}} \;  \frac{1}{2M_{\textrm{R}}} \textrm{Tr}(\textrm{Toep}(\mathbf{u}))+ \frac{1}{2} \textrm{Tr}(\mathbf{T}) \\ &\; \textrm{s.t.}
    \begin{bmatrix}
    \textrm{Toep}(\mathbf{u}) & \mathbf{Z} \\
    \mathbf{Z}^{H} & \mathbf{T}
    \end{bmatrix} \succeq 0,
    \end{split}
\end{equation}
where $\textrm{Toep}(\mathbf{u})$ denotes a Hermitian Toeplitz matrix whose first column equals $\mathbf{u}$.
With the ANM denoising theorem studied in~\cite{7313018}, an equation that estimates $\mathbf{Z}$ from $\mathbf{G}$ can be given by
\begin{equation}\label{final}   
    \hat{\mathbf{Z}}=\argmin_{\bar{\mathbf{Z}}} \;  \tau \lVert \bar{\mathbf{Z}} \rVert_{\mathcal{A}} + \frac{1}{2} \lVert \mathbf{G}-\mathbf{W}^{H}\bar{\mathbf{Z}} \rVert_{\textrm{F}}^{2},
\end{equation}
where $\hat{\mathbf{Z}}$ and $\bar{\mathbf{Z}}$ respectively denote the estimate of $\mathbf{Z}$ and the variable for estimation of $\mathbf{Z}$. 
A regularization parameter $\tau$ is set as in~\cite{7313018}: 
\begin{equation}\label{tau}   
    \begin{split}
    \tau=\frac{\sigma}{\sqrt{D}}&\Big(1+\frac{1}{\log M_{\textrm{R}}} \Big)^{\frac{1}{2}} \Big( M_{\textrm{B}}M_{\textrm{U}} + \log(\alpha M_{\textrm{B}}M_{\textrm{U}}) +  \\  &\sqrt{2M_{\textrm{B}}M_{\textrm{U}} \log(\alpha M_{\textrm{B}}M_{\textrm{U}})} + \sqrt{\frac{\pi M_{\textrm{B}}M_{\textrm{U}}}{2}} +1  \Big)^{\frac{1}{2}},
    \end{split}
\end{equation}
where $\alpha=8\pi M_{\textrm{R}} \log M_{\textrm{R}}$.
(\ref{final}) can be fully unfolded as
\begin{equation}\label{final2}   
    \begin{split}
    &\{\hat{\mathbf{u}},\hat{\mathbf{T}},\hat{\mathbf{Z}} \}=\\ & \argmin_{\mathbf{u},\mathbf{T},\bar{\mathbf{Z}}} \;  \frac{\tau}{2M_{\textrm{R}}} \textrm{Tr}(\textrm{Toep}(\mathbf{u}))+ \frac{\tau}{2} \textrm{Tr}(\mathbf{T}) + \frac{1}{2} \lVert \mathbf{G}-\mathbf{W}^{H}\bar{\mathbf{Z}} \rVert_{\textrm{F}}^{2} \\ &\; \textrm{s.t.}
    \begin{bmatrix}
    \textrm{Toep}(\mathbf{u}) & \bar{\mathbf{Z}} \\
    \bar{\mathbf{Z}}^{H} & \mathbf{T}
    \end{bmatrix} \succeq 0. 
    \end{split}
\end{equation}
Finally, $\mathbf{H}_{\textrm{eff}}$ can be approximated as $\hat{\mathbf{Z}}^{H}$. 
The proposed ANM-based low-overhead channel estimation for RIS-aided MIMO systems can be summarized as Fig.~\ref{Fig3}.

\section{Simulation Results and Discussions}\label{simulation}
In this section, the proposed algorithm is compared with \cite{9354904} and \cite{he2021channel}, and LS is added as a benchmark if $B=M_{\textrm{R}}$.
Note that \cite{9354904} and \cite{he2021channel} also employ the training beamwidth adaptation in this simulation. 
The algorithm in \cite{9103231} is excluded from the comparison since the computation time becomes excessively high when reforming the algorithm to work under RIS-aided MIMO systems.
For simulation, $M_{\textrm{B}}$, $M_{\textrm{R}}$, and $M_{\textrm{U}}$ are respectively set to 8, 32, and 4, and $N_{\textrm{B}}$ and $N_{\textrm{U}}$ are set to 4 and 2 so that there are 4 beam training per frame. 
$\alpha^{l}_{\textrm{BR}} \sim \mathcal{CN}(0,1)$ for $l=1,\ldots,L_{\textrm{BR}}$ and $\alpha^{l}_{\textrm{RU}} \sim \mathcal{CN}(0,1)$ for $l=1,\ldots,L_{\textrm{RU}}$.  
All AoAs and AoDs such as $\phi_{\textrm{BR}}^{l}$, $\theta_{\textrm{BR}}^{l}$, $\phi_{\textrm{RU}}^{l}$, and $\theta_{\textrm{RU}}^{l}$ are chosen randomly between $\left[30^{\circ},150^{\circ}\right]$.
The SNR is defined as
\begin{equation}\label{SNR}
    \textrm{SNR}=10 \log_{10} \frac{\left| \left( \sum_{l=1}^{L_{\textrm{RU}}} \alpha_{\textrm{RU}}^{l} \right) \left( \sum_{l=1}^{L_{\textrm{BR}}} \alpha_{\textrm{BR}}^{l} \right) \right|^{2}}{\sigma^{2}}  (\textrm{dB}).
\end{equation}
The SNR defined in~(\ref{SNR}) is a ratio of signal power and noise power when $M_{\textrm{B}}=M_{\textrm{R}}=M_{\textrm{U}}=1$. Note that the defined SNR does not depend on precoding matrix, combining matrix, and RIS control matrix. 
The NMSE is defined as
\begin{equation}
    \textrm{NMSE}=\frac{1}{Q}\sum_{q=1}^{Q} \frac{\lVert \hat{\mathbf{H}}^{q}_{\textrm{eff}} - \mathbf{H}^{q}_{\textrm{eff}} \rVert^{2}_{\textrm{F}}}{\lVert \mathbf{H}^{q}_{\textrm{eff}} \rVert^{2}_{\textrm{F}}},        
\end{equation}
where $Q$ is the number of Monte Carlo trials for NMSE calculation and is set to 300.
$\hat{\mathbf{H}}^{q}_{\textrm{eff}}$ and $\mathbf{H}^{q}_{\textrm{eff}}$ respectively denote the estimated effective channel and the actual effective channel on the $q$-th Monte Carlo trial.

\begin{figure}[!t]
    \includegraphics[width=1\columnwidth]{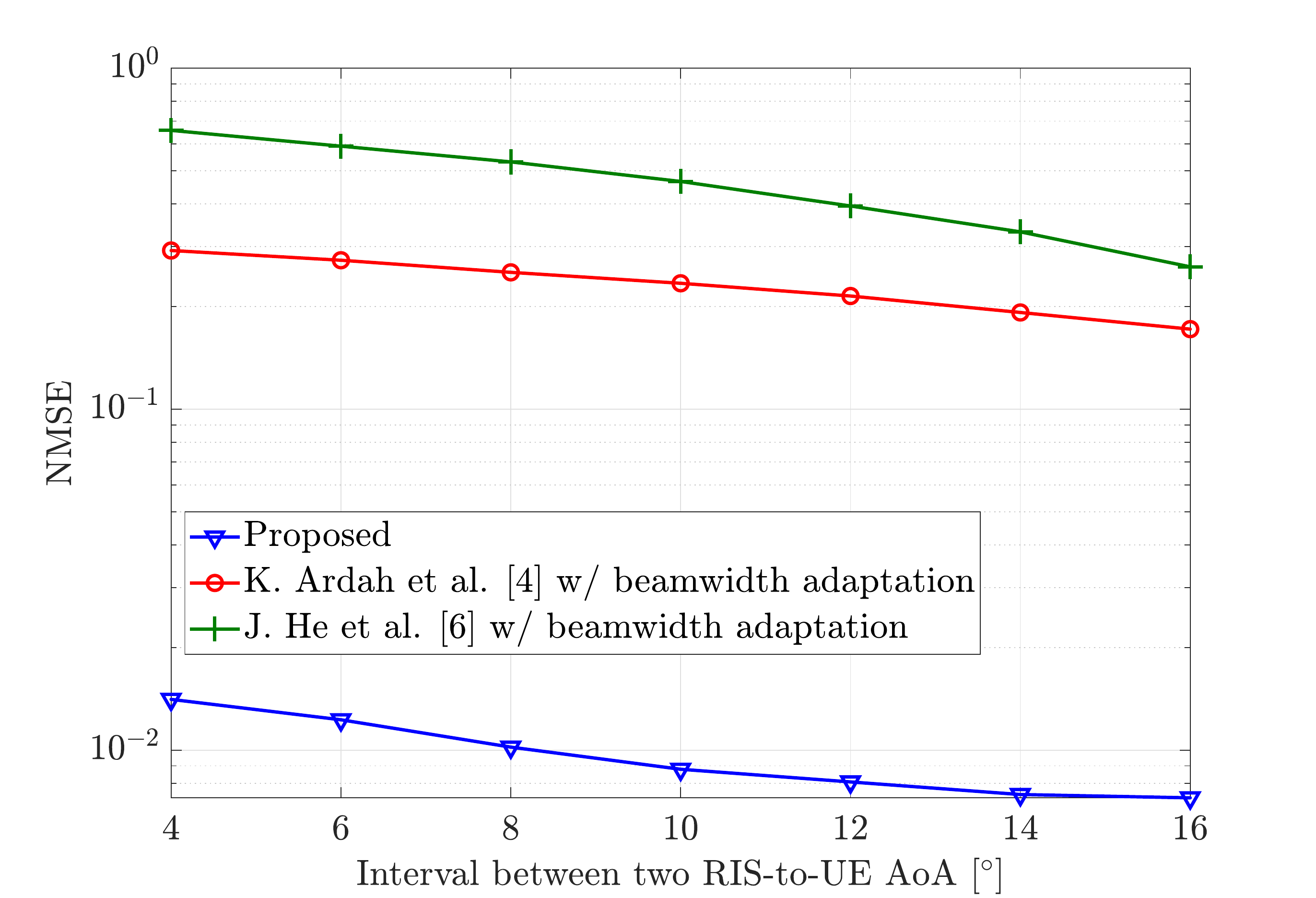}
    \caption{NMSE versus interval between two RIS-to-UE AoAs. The SNR is set to 0 dB. $L_{\textrm{BR}}=1$, $L_{\textrm{RU}}=2$, and $B=16$.}
    \label{NMSEDiff}
\end{figure}

Fig.~\ref{NMSEDiff} shows the NMSE versus interval between two RIS-to-UE AoAs when $L_{\textrm{BR}}=1$, $L_{\textrm{RU}}=2$, and $B=16$. The SNR is set to 0 dB.
Considering that AoDs and AoAs are uniformly distributed in all angles~\cite{7501500}, there is a possibility that AoDs and AoAs of signal paths are closely separated.
In \cite{9354904} and \cite{he2021channel}, AoD/AoA estimation can be inaccurate when AoDs and AoAs are closely separated, where inaccurate AoD/AoA estimation leads to channel estimation failure.
To show the correlation between the channel estimation accuracy and the separation between AoAs or AoDs, we analyze NMSE with respect to the interval between two RIS-to-UE AoAs. 
In Fig.~\ref{NMSEDiff}, the NMSE of \cite{9354904} and \cite{he2021channel} are high when the two RIS-to-UE AoAs are closely separated as the estimation on RIS-to-UE AoAs fails.
On the other hand, the NMSE of the proposed algorithm remains relatively low even when the interval between two RIS-to-UE AoAs is close.
When the interval between two RIS-to-UE AoAs is $4^{\circ}$, the NMSE of proposed algorithm, \cite{9354904}, and \cite{he2021channel} are $0.014$, $0.29$, and $0.66$ respectively.

\begin{figure}[!t]
    \includegraphics[width=1\columnwidth]{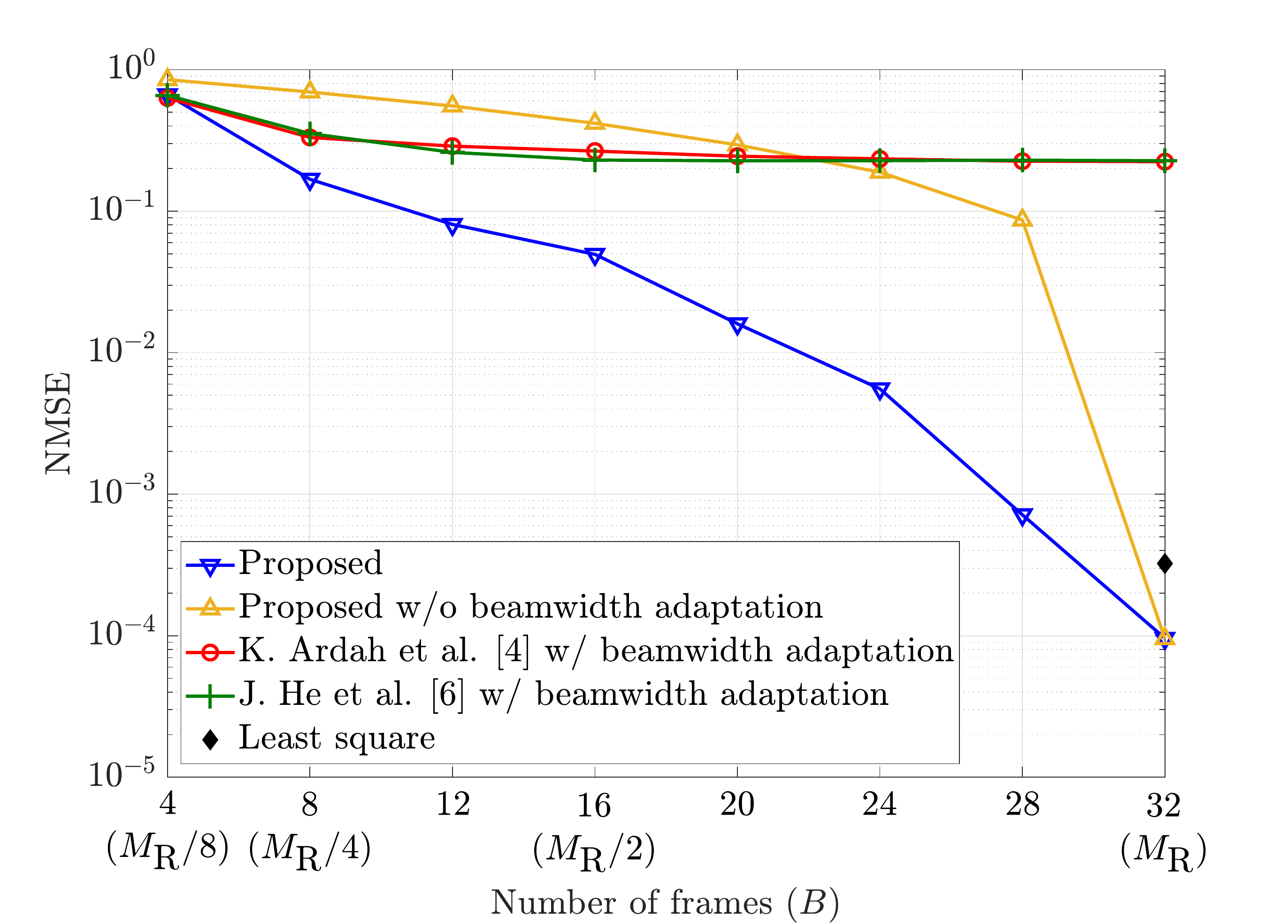}
    \caption{NMSE versus number of frames. The SNR is set to 0 dB. $L_{\textrm{BR}}=2$ and $L_{\textrm{RU}}=2$. The number of frames equals to the number of beams created by RIS during beam training.}
    \label{NMSERISTrain}
\end{figure}

Fig.~\ref{NMSERISTrain} shows the NMSE versus number of beam training when $L_{\textrm{BR}}=2$ and $L_{\textrm{RU}}=2$. The SNR is set to 0 dB. To show that the training beamwidth adaptation ensures robust channel estimation when there is less beam training, the NMSE of the proposed algorithm without training beamwidth adaptation is also presented in Fig.~\ref{NMSERISTrain}.
Since the deficiency of beam training may cause the erroneous multipath signal reception, the proposed algorithm without training beamwidth adaptation shows high NMSE when $B<M_{\textrm{R}}$.
On the other hand, the NMSE of the proposed algorithm is lower than those of other algorithms at every $B$.
When $B$ reaches $M_{\textrm{R}}$, there is no need for broadening beamwidth so that the NMSE of the proposed algorithm without training beamwidth adaptation becomes equivalent to the NMSE of the proposed algorithm.
% As shown in Fig.~\ref{NMSEDiff}, the estimation of \cite{9354904} and \cite{he2021channel} fail when AoDs or AoAs are closely separated.
Since the channel estimation is inaccurate when AoDs or AoAs are closely separated, the NMSE of \cite{9354904} and \cite{he2021channel} remains high even when sufficient beam training is performed. 

\begin{figure}[!t]
    \includegraphics[width=1\columnwidth]{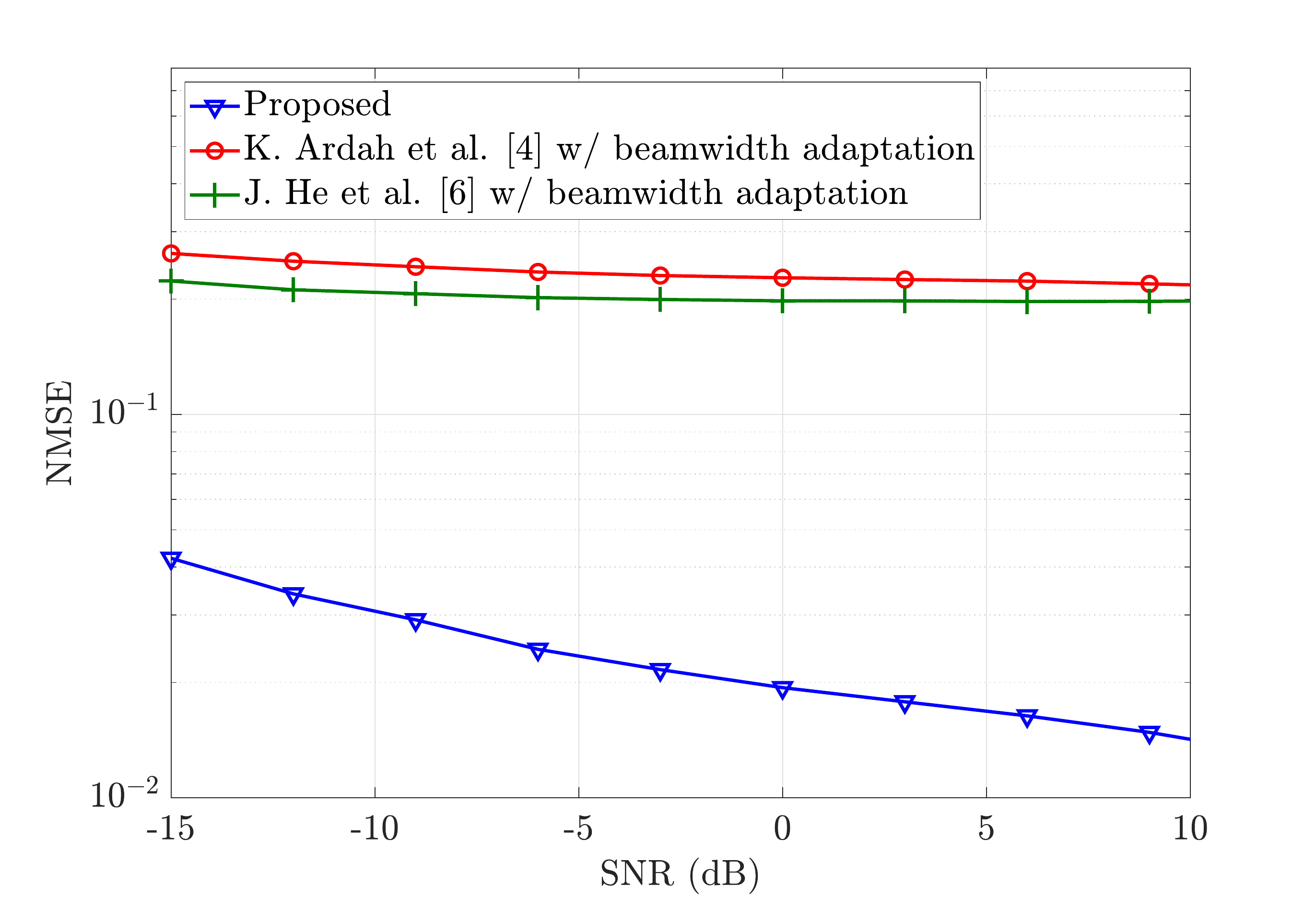}
    \caption{NMSE versus SNR. $L_{\textrm{BR}}=1$, $L_{\textrm{RU}}=2$, and $B=16$.}
    \label{NMSESNR}
\end{figure}

Fig.~\ref{NMSESNR} shows the NMSE versus SNR when $L_{\textrm{BR}}=2$, $L_{\textrm{RU}}=2$, and $B=16$.
The NMSE of the proposed algorithm is lower than those of other algorithms at every SNR.  
On the other hand, the NMSE of \cite{9354904} and \cite{he2021channel} do not improve with the increase of SNR and remains relatively high. 
For the same reason as previous results, this is because the estimation of \cite{9354904} and \cite{he2021channel} fails when AoDs or AoAs are closely separated, and this failure is independent of the SNR.

From Fig.~\ref{NMSERISTrain} and Fig.~\ref{NMSESNR}, we can conclude that the proposed algorithm shows the best channel estimation accuracy when the identical number of beam training and SNR are given. 
As shown in Fig.~\ref{NMSEDiff}, the superiority of the proposed algorithm is attributed to the robustness against close separation between AoDs and AoAs.
Also, Fig.~\ref{NMSERISTrain} supports that the training beamwidth control makes channel estimation accurate when there is less beam training.

However, the channel estimation becomes inaccurate as the number of active RIS antennas decreases, where this inaccuracy is induced by the decrease of beam gain and the decrease of the number of detectable signal paths.
Thus, the idea to maintain high channel estimation accuracy while extremely lowering beam training overhead needs to be discussed in further work, as well as the analysis on the relationship between channel estimation accuracy and the number of active RIS antennas.

\section{Conclusions}
In this paper, we propose a low-overhead channel estimation algorithm for RIS-aided MIMO systems.
When there is less beam training, some multipath signals may not be received, and this causes the channel estimation failure.
To address this issue, the beamwidth of RIS is adaptively widened so that the beamwidth is inversely proportional to the number of beams created by RIS.
The atomic norm of the effective channel for RIS-aided MIMO systems is defined, where defining the atomic norm requires the compilation of pilot signals received during beam training.
The effective channel is estimated by solving the SDP that represents the atomic norm.
Simulation results show that the proposed channel estimation algorithm has the lowest NMSE among other algorithms for RIS-aided MIMO systems when the identical number of beam training and SNR are given.

\ifCLASSOPTIONcaptionsoff
  \newpage
\fi

\bibliographystyle{ieeetr}
\bibliography{reference}
\end{document}